\documentclass{iaus}
\usepackage{graphicx}
\begin{document}

\title[NIR spectroscopy of Westerlund 1] 
{NIR spectroscopy of the most massive open cluster in the Galaxy: Westerlund 1}

\author[Mengel and Tacconi-Garman]   
{S. Mengel$^1$, L.E. Tacconi-Garman$^1$}

\affiliation{$^1$ESO, Karl-Schwarzschild-Str. 2, 85748 Garching, Germany
\break email: smengel@eso.org, ltacconi@eso.org\\}

\pubyear{2007}
\volume{246}  
\jname{Dynamical Evolution of Dense Stellar Systems}
\editors{E. Vesperini, M. Giersz, A. Sills, eds.}


\maketitle

\begin{abstract}
Using ISAAC/VLT, we have obtained individual spectra of all NIR-bright stars in the central
2'x2' of the cluster Westerlund 1 (Wd 1) with a resolution of R$\approx$9000 at a central wavelength of
2.30$\mu$m. This allowed us to determine radial velocities of ten post-main-sequence
stars, and from these values a velocity dispersion. Assuming virial equilibrium,
the dispersion of $\sigma$=8.4 km/s leads to a total dynamical cluster mass of
1.25$\times 10^5$M$_\odot$, comparable to the photometric mass of the cluster.
There is no extra-virial motion which would have to be interpreted as a signature
of cluster expansion or dissolution.

\end{abstract}

\firstsection 
\section{Observations and Results}
\begin{figure}
\begin{center}
\includegraphics[angle=270,width=10.5cm]{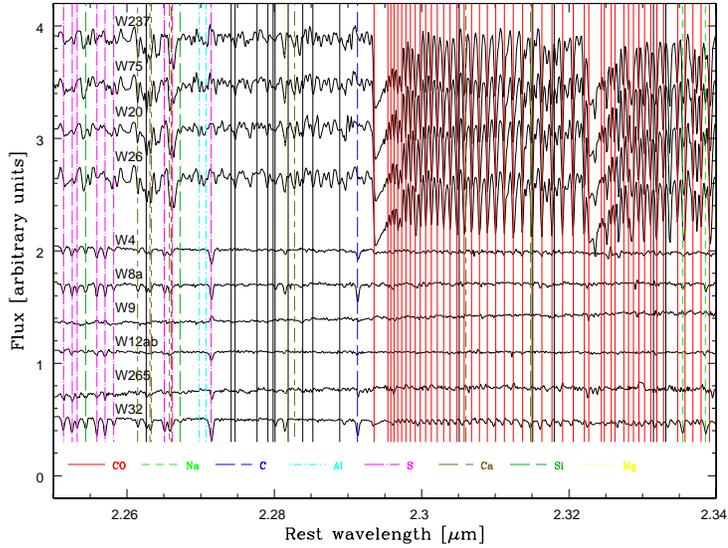}
\end{center}
\caption{Normalized, vertically offset spectra of all detected cluster members with absorption features 
in our observed wavelength regime. Top four are the RSGs, five of the six below are YHGs (W9 is a sgB[e] 
star).\label{spectra}}
\end{figure}

We used ISAAC/VLT in its highest spectral resolution mode (R$\approx$9000) to step the
2' long slit across the centre of the cluster for 2', in
steps of 0.3''. At each position, the integration time was 3 x 1.77s.
The central wavelength of 2.31$\mu$m was optimized for the analysis or red supergiant (RSG) spectra,
which show deep CO absorption features in this regime.
The data were sky subtracted, flat fielded, distortion corrected and wavelength calibrated using
standard {\sl iraf} routines and, where necessary, calibration observations.
Individual spectra were extracted from those slit positions where a star with photospheric
features in our wavelength range was included. After the extraction of the 1d spectrum, we corrected
for telluric absorption.
Fig. \ref{spectra} shows the ten spectra which were extracted from our scan, four red supergiants (RSGs),
five yellow hypergiants (YHGs), and one sgB[e] star.

We used the lines in these spectra to determine the radial
velocities and their dispersion, which is
$\sigma$=8.4 km/s. This, together with the half-light radius (estimated from archival 
NTT/SOFI images), assuming that the Virial
Theorem is valid, yields a dynamical cluster mass of 1.3$\times 10^5$M$_\odot$
(see also \cite{Mengel07}).

\begin{figure}
\begin{center}
\includegraphics[angle=0,width=6.5cm]{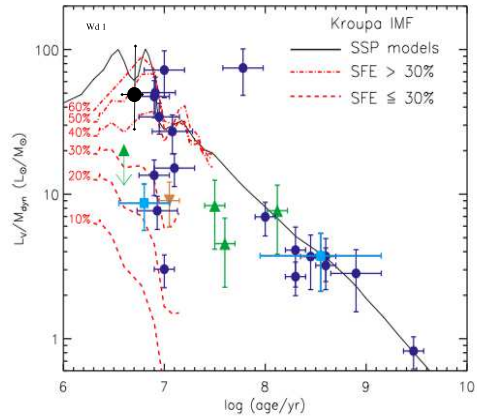}
\end{center}
\caption{L$_V$/M$_{dyn}$ as a function of age, for ``static'' clusters (solid)
and clusters adjusting dynamically after gas expulsion (from \cite{GoodwinBastian}), dashed.
Wd 1 (black dot) does not seem to be expanding dramatically.\label{GoodwinBastian}}
\end{figure}

Comparing the light-to-mass-ratio (V-band luminosity and dynamical mass) to that
expected for a cluster of this age and luminosity, and a Kroupa IMF, solar metallicity,
allows a conclusion regarding the dynamical state of the cluster: Even though it
could in principle also be an indication of an excess of low-mass stars (compared to
the Kroupa IMF), the more likely explanation for low L/M is that the high ``dynamical mass'' arises
from extra-virial motion, and hence an expanding cluster.
Fig. \ref{GoodwinBastian} shows a plot taken from \cite{GoodwinBastian}, where they
determine the velocity dispersion of the stars in their N-body simulated clusters and
plot the resulting L/M ratios as dashed lines. Clusters with a low star-formation efficiency
show large amounts of extra-virial motion, and do not survive as bound clusters.

The L/M ratio for Westerlund 1 is consistent with the expectation from the evolutionary
synthesis model, which means that the cluster may survive for
a long time, possibly several Gyr.

The most important caveat in our measurement is the low number of stars, which is why we
want to improve statistics by obtaining radial velocities also for the hot stars in the cluster, 
from absorption lines around Br$\gamma$.

\end{document}